# TURBULENT MICROSCALE FLOW FIELD PREDICTION IN POROUS MEDIA USING CONVOLUTIONAL NEURAL NETWORKS

**Vishal Srikanth[1], Ching-Wei Huang[1], Ryan Harradine[1], Andrey V. Kuznetsov[1*]**

[1]Department of Mechanical and Aerospace Engineering, North Carolina State University, Raleigh, NC 27695, USA

## ABSTRACT

Turbulence modeling in porous media can be greatly improved by combining high-resolution numerical methods with modern data-driven techniques. The development of accurate macroscale models (length scale greater than the pore size) will enable real-time systemic simulations of porous media flow. We consider the case of turbulent flow in homogeneous porous media, typically encountered in engineered porous media (heat exchangers, metamaterials, combustors, etc.). The underlying microscale flow field is inhomogeneous and determined by the geometry of the porous medium. Neural Networks are able to resolve the geometry-dependence and the non-linearity of porous media turbulent flow. We are proposing to separate the macroscale model into individual blocks that predict a unique aspect of the microscale flow, such as microscale spatial flow distribution and vortex dynamics. In the present work, we determine the feasibility of the prediction of the Reynolds-averaged microscale flow patterns by using Convolutional Neural Networks (CNN).

The porous medium is represented by using a square lattice arrangement of circular cylinder solid obstacles. The pore-scale Reynolds number of the flow is 300. The porosity of the porous medium is varied from 0.45 to 0.92 with 60 steps. The microscale flow field is simulated by using Large Eddy Simulation (LES) with a compact sixth-order finite difference method. We demonstrate satisfactory prediction of the microscale flow field using the CNN with a global error less than 10%. We vary the number of training samples to study the deterioration of the model accuracy. The CNN model offers a $O(10^6)$ speedup over LES with only 10% loss in accuracy.

## 1. INTRODUCTION

Microscale turbulent flow in porous media is highly nonlinear and inhomogeneous such that traditional macroscale modelling approaches cannot be both robust and accurate. In this context, microscale refers to a length scale that is smaller than the size of the pore. Therefore, classical porous medium models are limited to a particular geometric configuration. There is a growing interest in designing porous materials with targeted attributes for a range of applications. Porous heat exchangers can be designed to optimise transport to meet the growing demand in electronics cooling (Jiang *et al.* 2001; Vafai 2015). Porous metamaterials are used in flow control and stealth for aerospace applications (Repasky & Alexander 2019). There is also a need to improve macroscale models for accurate prediction of the microscale transport (Vafai *et al.* 2009). Macroscale models improve the prediction of forest fire spread at a nominal cost (Rehm & Evans 2011). They can predict the transmission of viruses through masks and filters (Mittal *et al.* 2020). A straightforward solution to simulate turbulent flow in these systems is to represent the solid and fluid phases explicitly. This is not practical for systemic simulations when the grid and time step size are taken into consideration. The microscale flow field is typically simulated using representative volumes to study the fundamental flow physics (de Lemos 2006).

The microscale flow field inside the porous medium is complex because of the solid matrix geometry. In spite of the geometry dependence, the fundamental physical qualities of the flow can be extracted from microscale simulations for the purpose of modelling. The characteristic length scale of turbulence can be determined by the size of the pore. This is supported by the Direct Numerical Simulations (DNS) of Jin *et al.* (2015) and Uth *et al.* (2016), which show that the size of the turbulence structures in a porous medium is of the order of the pore scale. The DNS studies in a dense porous medium by He *et al.* (2019) also showed that the turbulence integral length scale is limited to ~10% of the solid obstacle diameter. The turbulence time scales can be described from the flow physics, since the integral timescales of the turbulent structures are limited (He *et al.* 2018). The microscale flow distribution within a porous medium is known to possess the characteristics of both classical internal and external flows. There are distinct regions of vortex-dominated and boundary-layer-dominated flow. The unique features of porous media flow can be analysed qualitatively using DNS, such as the significant regions of negative TKE production that imply a local reversal in the TKE transport process. The tortuous path in porous media also selectively reduces the streamwise turbulence intensity at locations of localized adverse pressure gradient. The turbulence anisotropy is concentrated near the solid walls, and near-isotropic behaviour was observed in the bulk flow (Chu *et al.* 2018). As the Reynolds numbers increase, the large scale energetic eddies were observed to break down into smaller structures resulting in a microscopic flow that can be assumed locally isotropic. This finding is consistent across DNS studies (Chu *et al.* 2018; He *et al.* 2019) and PIV measurements for packed beds (Khayamyan *et al.* 2017; Nguyen *et al.* 2019; Patil & Liburdy 2015). These observations indicate that the solid obstacle geometry determines both the non-linearity and inhomogeneity in the microscale flow field. The different microscale transport processes can be expressed qualitatively using traditional models, but it is near impossible to predict them quantitatively for all geometries and Reynolds numbers. In the past, this limitation was overcome by introducing empirical constants, which requires expensive experiments (Nield & Bejan 2017).

*Corresponding Author: avkuznet@ncsu.edu

Data-driven methods show great promise in overcoming the limitations of porous media modelling by serving as an interface between microscale simulations and macroscale modelling. Neural networks are used in fluid mechanics for reduced-order modelling, analyse the flow and identify patterns, and even to discover governing equations (Duraisamy *et al.* 2019). Neural networks are conducive to the modelling of the material properties of composite domains (Chung *et al.* 2020; Rabbani & Babaei 2019; Wei *et al.* 2018). The advantage of using neural networks is the inclusion of morphological effects in the model with physics-based descriptors like the porosity (Wei *et al.* 2020). The neural network approach is a form of interpolation that takes advantage of the Universal Approximation Theorem to fit any non-linear function if the appropriate weights are discovered. In turbulence, neural networks have been used to model the flow around airfoils using a single layer of perceptrons or neurons (Zhu *et al.* 2019). For porous media flows, deep learning is required to fit the variation of the flow topology with the geometry of the solid obstacles. A Fully-connected Multi-Layer Perceptron network is a straightforward option, but it requires a considerable computation effort to train because all of the perceptrons are connected. Convolutional Neural Networks (CNN) are an efficient alternative for flow field prediction in porous media that breaks down the information into a feature map. CNNs have been used to upscale microscale flow to develop macroscale parameters with a comparable accuracy to the microscale simulations (Beck *et al.* 2019; Tahmasebi *et al.* 2020; Tembely *et al.* 2020; Vasilyeva *et al.* 2020). CNNs have been successfully used to predict laminar flow in random 2D porous media (Santos *et al.* 2020; Takbiri-Borujeni *et al.* 2020). However, a large number of samples have been used in the training by following a brute force approach.

Our approach is to develop a suite of neural network models that come together to predict the various aspects of the flow. Different neural network architectures are effective in modelling different types of systems. For example, the CNN models are capable of describing the spatial features of the flow that correspond to the given geometry. The Long Short-Term Memory (LSTM) Recurrent Neural Network architecture is effective in modelling the time dependent processes, such as turbulence dissipation or the vortex shedding. In this paper, we are interested in predicting the microscale turbulence flow field in porous media using CNNs for a systematic variation in the solid obstacle geometry. The objective of this paper is to systematically identify the number of samples of the microscale flow field that are required to predict the macroscale flow field with low error. Typically, a large number of samples are used to train neural networks in practice. We want to inform researchers of the minimum number of samples that would be needed for the geometry and flow parameters used in porous media modelling, so that future models can be more efficient. Since neural networks are an interpolation method, a systematic training is required to consider all of the regimes of flow that are observed. A large number of samples will not increase the robustness if the samples represent only a single regime. We have chosen to limit the scope to the variation of porosity for circular cylinder solid obstacles. The resulting predictive model yields an accurate solution at a fraction of the time it takes to simulate the microscale flow field using DNS. We also demonstrate the reduction of the modelling error with the increase in the training samples for the change in porosity to serve as a benchmark for future work.

## 2. SOLUTION METHODOLOGY

Data-driven methods like CNN models require training, testing and validation datasets as a precursor. The datasets are obtained from numerical simulations of the microscale flow field in porous media. We consider the case of homogeneous porous media consisting of a square lattice arrangement of circular cylinder solid obstacles (see Fig. 1(a)). The solution methodology consists of two components: (1) microscale flow simulation and (2) neural network training.

(a) (b)

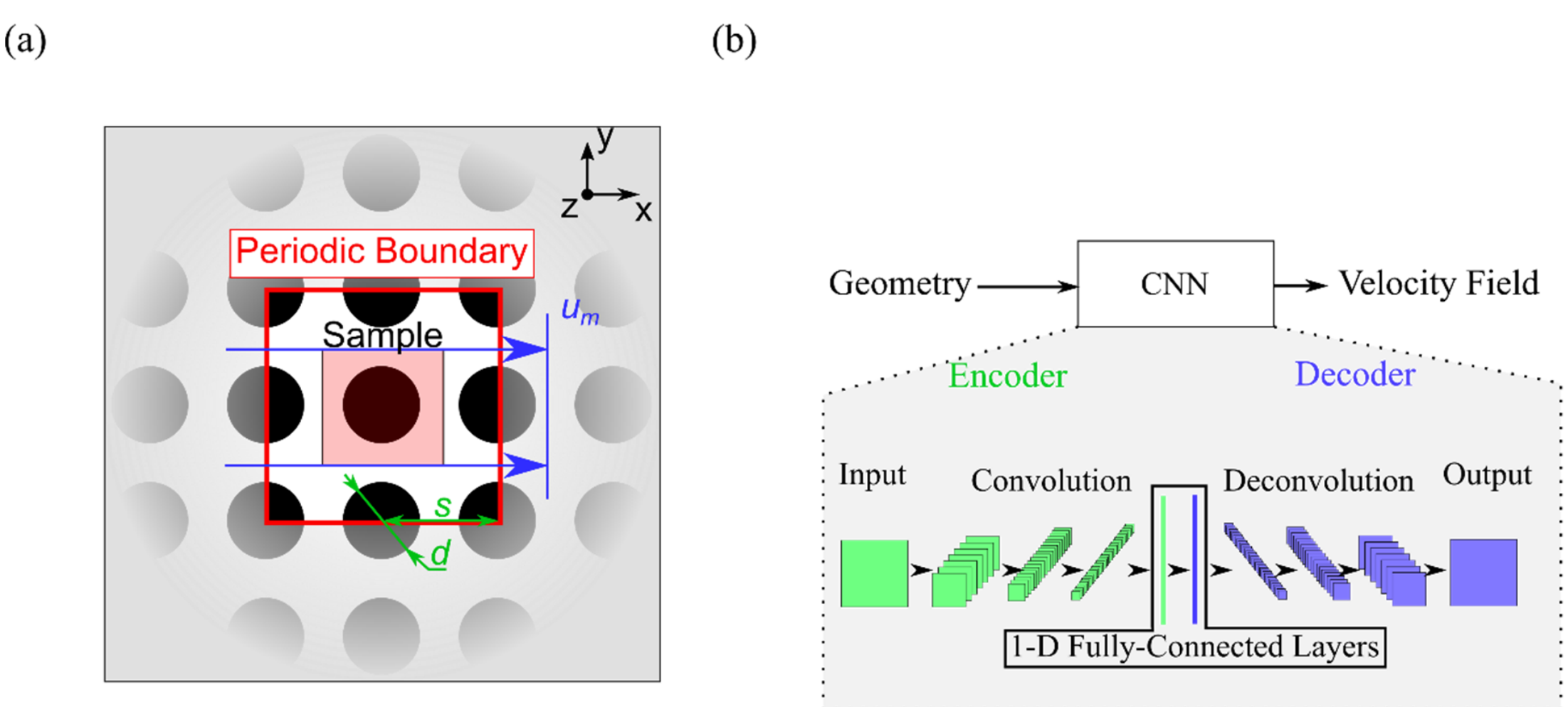

**Fig. 1** (a) Representative Elementary Volume (REV) used to simulate microscale turbulent flow. The sampling domain used to train the neural network is shown as a pink box. (b) The CNN architecture used to encode and decode the microscale flow field.

**2.1 Microscale flow simulation** The turbulent flow inside the Representative Elementary Volume (REV) domain shown in Fig. 1(a) is simulated by using the Fortran code, Incompact3d (Laizet & Lamballais 2009). The Incompact3d code has been validated with experimental data for a variety of canonical flows in the original article. Since the Neural Network model is the primary focus of this paper, experimental validation of the porous medium flow is not undertaken. Incompact3d makes use of the Finite Difference Method with structured grids of equidistant spacing to fill the entire REV-T. The Immersed Boundary Method (IBM) is used to account for the solid obstacles that are inside the computational grid. The Navier-Stokes equations written in equations (1)-(2) are solved using the Finite Difference Method (FDM). The velocities $u_i$ and the pressure $p$ are the primary variables. The Reynolds number is specified by setting the kinematic viscosity $\nu = 1/Re_p$. The Reynolds number used in this work is the pore scale Reynolds number ($Re_p$) which is calculated using equation 3. The characteristic velocity $u_m$ is the volume average of the $x$- velocity over the REV. The pore size $s$ is calculated from the solid obstacle diameter $d$ and the porosity φ. A momentum source term $f_i$ is used to represent the solid obstacles using the IBM. The direct forcing method is used such that the no-slip condition is realized at the immersed walls.

$$\frac{\partial u_j}{\partial x_j} = 0 \tag{1}$$

$$\frac{\partial u_i}{\partial t} = -\frac{\partial p}{\partial x_i} - \frac{1}{2}\left(\frac{\partial u_i u_j}{\partial x_j} + u_j \frac{\partial u_i}{\partial x_j}\right) + \nu \frac{\partial^2 u_i}{\partial x_i \partial x_j} + f_i + g_i \tag{2}$$

$$Re_p = \frac{u_m d}{\nu} \tag{3}$$

The spatial derivatives are approximated using a compact, sixth-order accurate Hermitian scheme. The location of the pressure variable is staggered by a half-mesh distance from the location of the velocity variable. The pressure Poisson equation is solved in Fourier space with the help of Fast Fourier Transform (FFT) routines. The governing equations are solved in a segregated manner using a three-step Fractional Step Method (FSM). A second-order accurate, explicit Adams-Bashforth method is used for time advancement. The momentum source term $g_i$ is adjusted every time-step to maintain a constant flow rate at the periodic boundaries. The method is implemented in the code in the non-dimensional form. A time step of 0.0002 non-dimensional units and a grid step of 0.03 non-dimensional units are used (based on Srikanth *et al.* (2020)). The REV consists of a 2x2 matrix of solid obstacles to include the influence of the phase difference in the vortex shedding process behind each solid obstacle on the mean flow variables. All of the simulations that are presented in this work are three-dimensional. Flow statistics have been averaged over 100 flow cycles for the sample volume. The simulations have been run on the North Carolina State University Linux Cluster. Suggestive computation time for each case is 7,000 CPU-Hours (1 CPU-Hour = Computation Time in Hours for a single CPU).

**2.2 Neural network training** The purpose of the neural network is to interpret geometric data and use it to predict the velocity field data (Fig. 1(b)). The network architecture is similar to that of Guo *et al.* (2016). The input is resampled into a 256x256 tensor. The encoding is performed with two filter layers: 256 filters with a 16x16 kernel (256@16x16) and 512 filters with an 8x8 kernel (512@8x8). A 2x2x512 tensor output is reshaped into a 1x1x2048 flat tensor. The flat tensor is connected to a second of the same size along with the ratio *d*/*s* as a secondary input to make the model physics-informed. The secondary input is a feature map of 1,024 identical values equal to the ratio, which ensures proper weighting of the physical parameter. The second flat tensor is reshaped to a 2x2x512 tensor, followed by the deconvolution filters: 512@8x8, 256@4x4, 32@2x2, and 1@1x1. The ReLU activation function is used throughout the network to take advantage of its ability to extract features. The Mean Squared Error loss function is used to measure the fit. The Adam optimization algorithm is used to iterate the system. The neural network was built and trained using the Tensorflow library for Python. The training was performed on a workstation using GPU acceleration with the NVIDIA CUDA library.

## 3. RESULTS AND DISCUSSION

Once the neural network is built, the weights have to be determined using a training dataset by applying an iterative method. This step will complete the model development. The physics-based neural network used in this work requires the porosity as a geometric input. The pore size *s* is set to vary uniformly in the range 1.2 to 3.0 with steps of 0.03. This results in quadratic steps for the porosity in the range of 0.454 to 0.913. The microscale geometry of the porous medium is given as the input to the neural network model. The geometry was initially represented using the distribution of the volume fraction of the solid phase. The discontinuity in the volume fraction function at the solid obstacle surface makes it less desirable for the numerical method used to train the neural network. To improve the accuracy, an exponential Signed Distance Function (SDF) is used to represent the geometry. The SDF assumes negative values in the solid phase, positive values in the fluid phase, and zero at the solid obstacle surface. The SDF for all of the samples is plotted in Fig. 2(a). The exponential function is chosen so that the solution is weighted at the solid obstacle surface where steep gradients are expected.

In this paper, the *x*- velocity distribution corresponding to the geometry input is the output of the neural network that it is trained to predict. The CNN model can also be used to predict any of the flow variables including the velocity, pressure, turbulence kinetic energy, and turbulence dissipation rate. Put together these variables can serve as closure for a non-linear macroscale turbulence model. The *x*- velocity distribution for all of the cases is simulated by using LES using the numerical method described in 2.1. The LES solution is plotted in Fig. 2(b). The 60 sample cases are split between training and testing in the ratio 75:25 (training: testing). The test cases are uniformly spaced across the entire range of pore sizes. The neural network model is trained using the training dataset until the loss function drops below a threshold value of $2 \times 10^{-4}$. Further training is observed to increase the loss function value by two orders of magnitude. The predicted *x*- velocity distribution from the CNN model is plotted in Fig. 2(c). The distribution is predicted at all of the sample data points, but every 4$^{th}$ case corresponds to a test case. The test cases show greater error since the model has not been trained to predict their distribution. However, it is apparent from visual inspection that the CNN model predicts the features of the flow and the magnitude of the *x*- velocity accurately for all of the cases when compared to LES. The prediction of features in the data is one of the strengths of the CNN architecture. It is the reason why the CNN model is used in this work to predict the microscale flow distribution for different geometries.

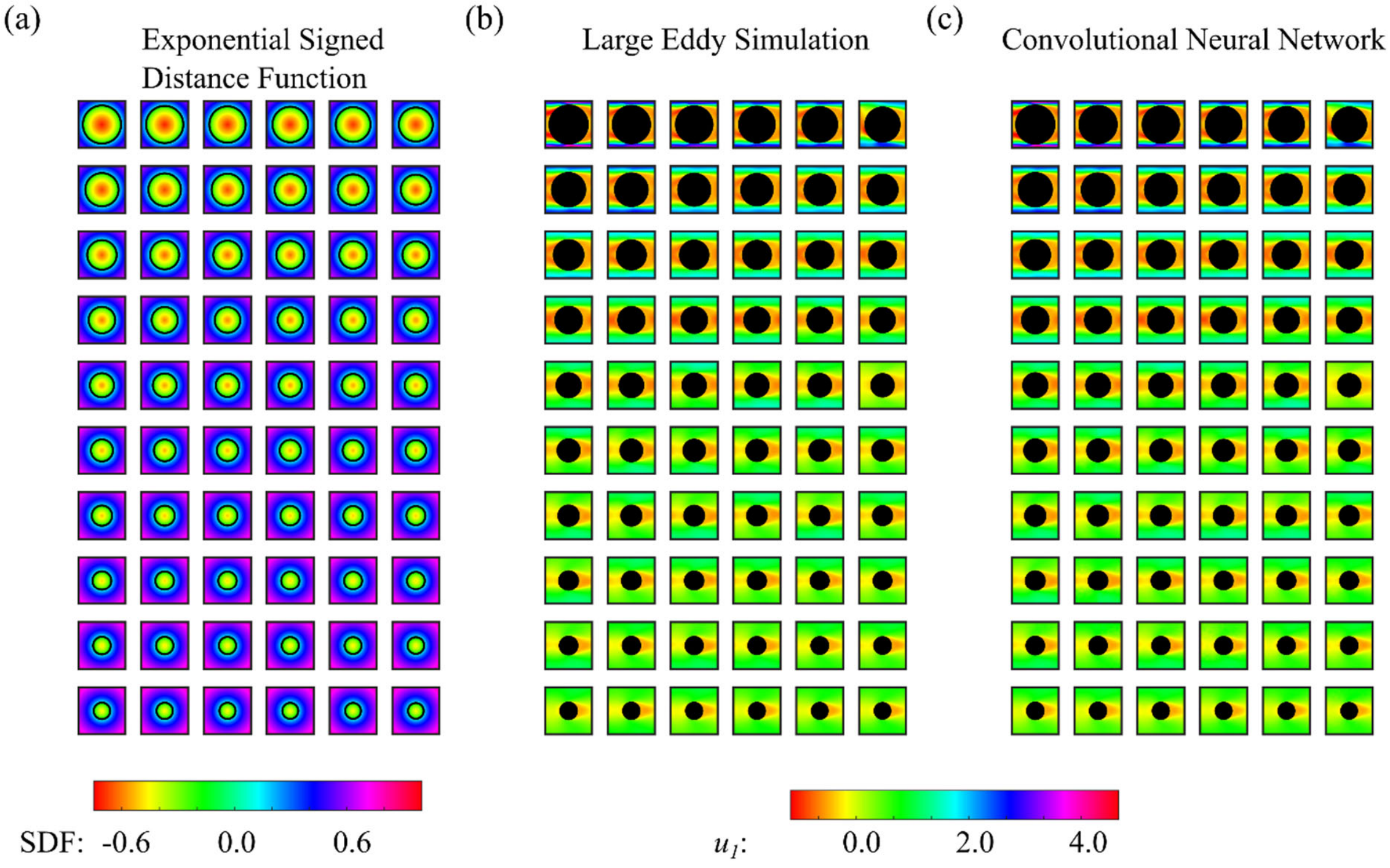

**Fig. 2** The dataset that is used to model the microscale *x*- velocity field in the porous medium. (a) The exponential SDF is the input to the model. The black circles show the solid obstacle surface. (b) The sample dataset for the training and testing of the CNN model obtained from LES. (c) The prediction dataset obtained from the CNN model.

Qualitatively, the CNN model is able to predict the microscale flow field. It should be noted that the model was trained by using a solution field that is not smooth. The model performs adequately even in the presence of gradient discontinuities at the solid obstacle surface. However, it can be a source of quantitative mismatch. We examine the LES and CNN *x*- velocity distributions for the case of $\varphi = 0.731$ to determine the microscale distribution of the prediction error (Fig. 3). The CNN prediction error is calculated as the difference between the CNN and LES solutions and normalized using the LES solution. While there is no discernable difference between the LES and CNN solution fields, the CNN solution has a small amount of noise (< 1%) visible at locations of steep gradients.

The distribution of the CNN prediction error (Fig. 3(c)) shows high magnitude of error in the vortex wake region behind the solid obstacle. The error is localized at the boundary of the vortex where a strong shear layer is present, which is characterized by a steep gradient in the *x*- velocity. The other considerations are the nonlinear changes in the vortex diameter (normalized by the pore size) and the maximum *x*- velocity inside the porous medium with the porosity. The error is also localized at the solid obstacle surface (not visible in Fig. 3), which stems from the steep gradient in the boundary layer. These sources contribute a Root Mean Squared Error (RMSE) of 13.91% for this case, even though the predicted and simulated solutions appear identical. The choice of the error function is crucial to the determination of the accuracy of the model. Even with the large RMSE microscale error, the CNN model predicts the mean macroscale *x*- velocity with an error of 0.26%. The exact location and size of the vortex are not important for the prediction of the macroscale turbulence variables. However, the accurate prediction of the boundary layer is important for modeling the microscale drag. The application of CNN to a non-smooth velocity data matrix will not yield an effective model for the boundary layer and drag predictions.

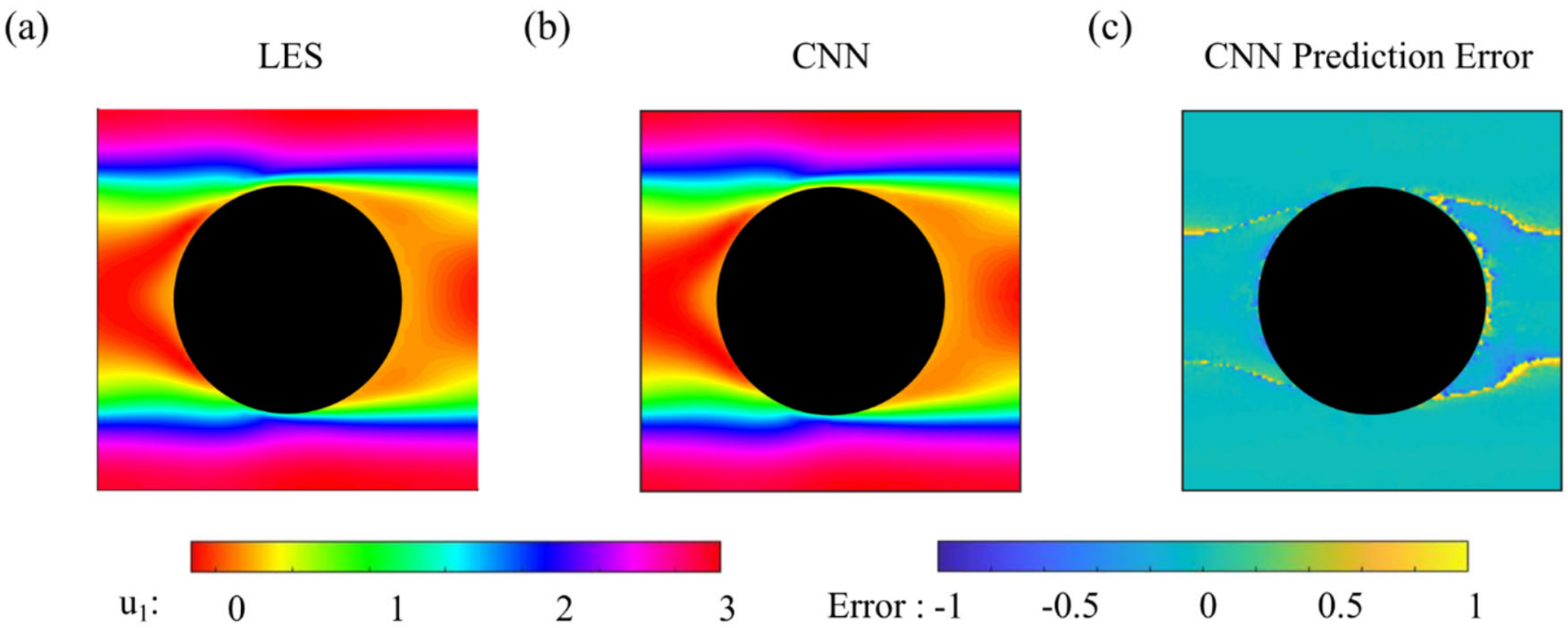

**Fig. 3** (a) LES and (b) CNN (60 samples) predictions for φ = 0.731 and (c) the CNN prediction error at each data point in space.

Next, we look at the number of samples that are used to train the CNN model. The porosity is the only parameter that is varied in this work. For neural network models to become viable for porous media turbulence modeling, the model should be trained by a systematic variation of several sensitive parameters. Therefore, 60 samples is an excessive number to train the model for a single parameter while keeping the solid obstacle shape, arrangement, and Reynolds number unchanged. The variation of the prediction error distribution with the number of samples is shown in Fig. 4. The superior performance of the model trained with 60 samples is reasonably matched when 30 samples are used. In the case of 30 samples, pixelation is observed in the error distribution that comes from the autoencoder process of the CNN model. A marginally greater mismatch in the boundary layer and the vortex wake location are observed. However, the error continues to be localized and marginal. With 30 samples, the RMSE error is 25.6% and the error in the macroscale velocity prediction is 0.55%. When the number of samples is reduced to 15 samples, the error grows further, but it is still localized. The RMSE error is 36.76% and the error in the macroscale velocity prediction is 2.11%. When the LES and CNN solutions were inspected (not shown here), the topological features of the flow such as the locations of the vortices, the flow separation points, and the stagnation point are predicted accurately.

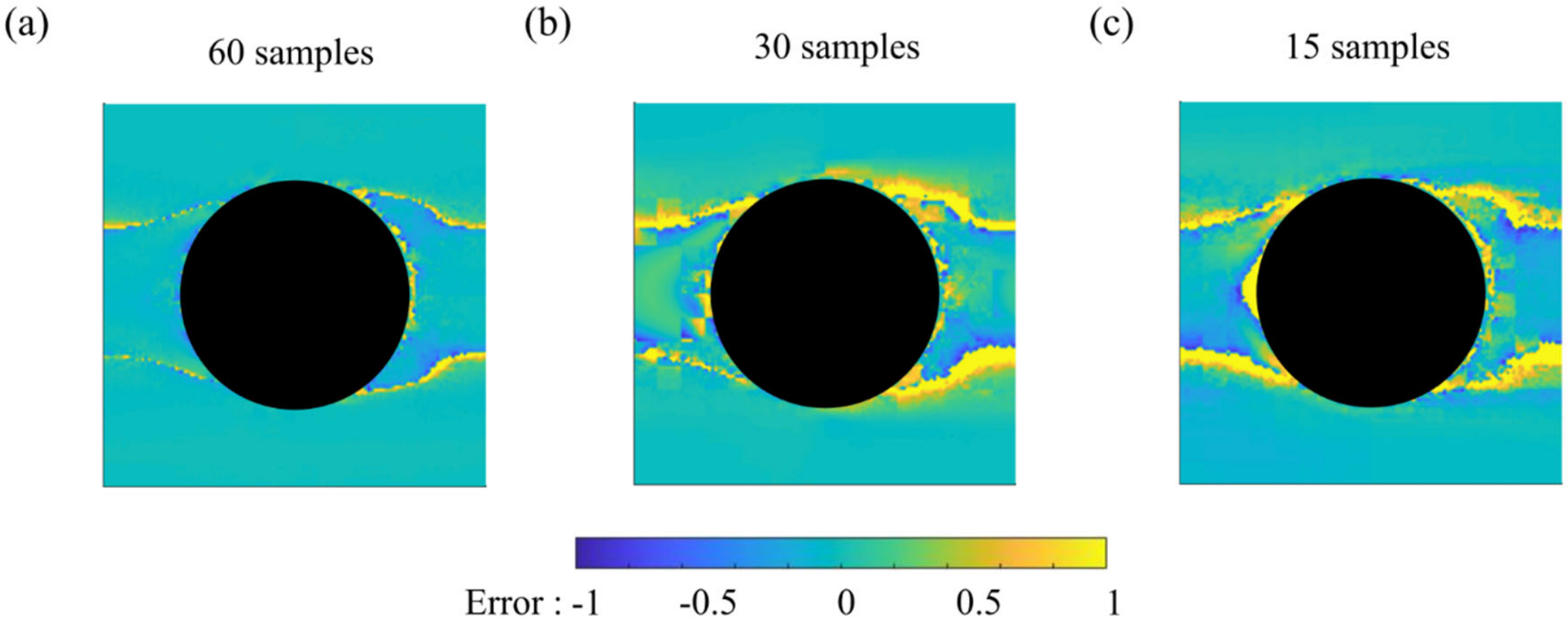

**Fig. 4** CNN prediction error for (a) 60, (b) 30, and (c) 15 input samples for φ = 0.811. The samples are split 75% training and 25% testing.

The CNN model predicts the flow separation and stagnation points accurately with only 15 samples (11 training cases). This result is expected given that CNNs are developed for computer vision, where the identification of topological features is of paramount importance. A remarkable observation in the present work is the ability of the CNN model to predict the mean maximum *x*- velocity, which is a local parameter that does not directly benefit from data convolution. The predicted variation in the maximum *x*- velocity with porosity is plotted in Fig. 5(a). The variation of the global error over all the simulated cases with the number of samples is shown in Fig. 5(b). The three types of error plotted are: (1) the normalized residual, (2) the RMSE, (3) the prediction error for the mean macroscale *x*- velocity. The normalized residual is calculated as the Euclidean norm of the difference between the LES and CNN data divided by the Euclidean norm of the LES data. Since macroscale turbulence models are of interest, the macroscale error is the most relevant metric here. The macroscale error is consistently below 10% for all of the simulated cases even when only 8 samples are used to train the CNN model. The use of a large number of samples for the single parameter (porosity) is not required to overcome the accuracy limitations of the traditional RANS models for porous media turbulent flow.

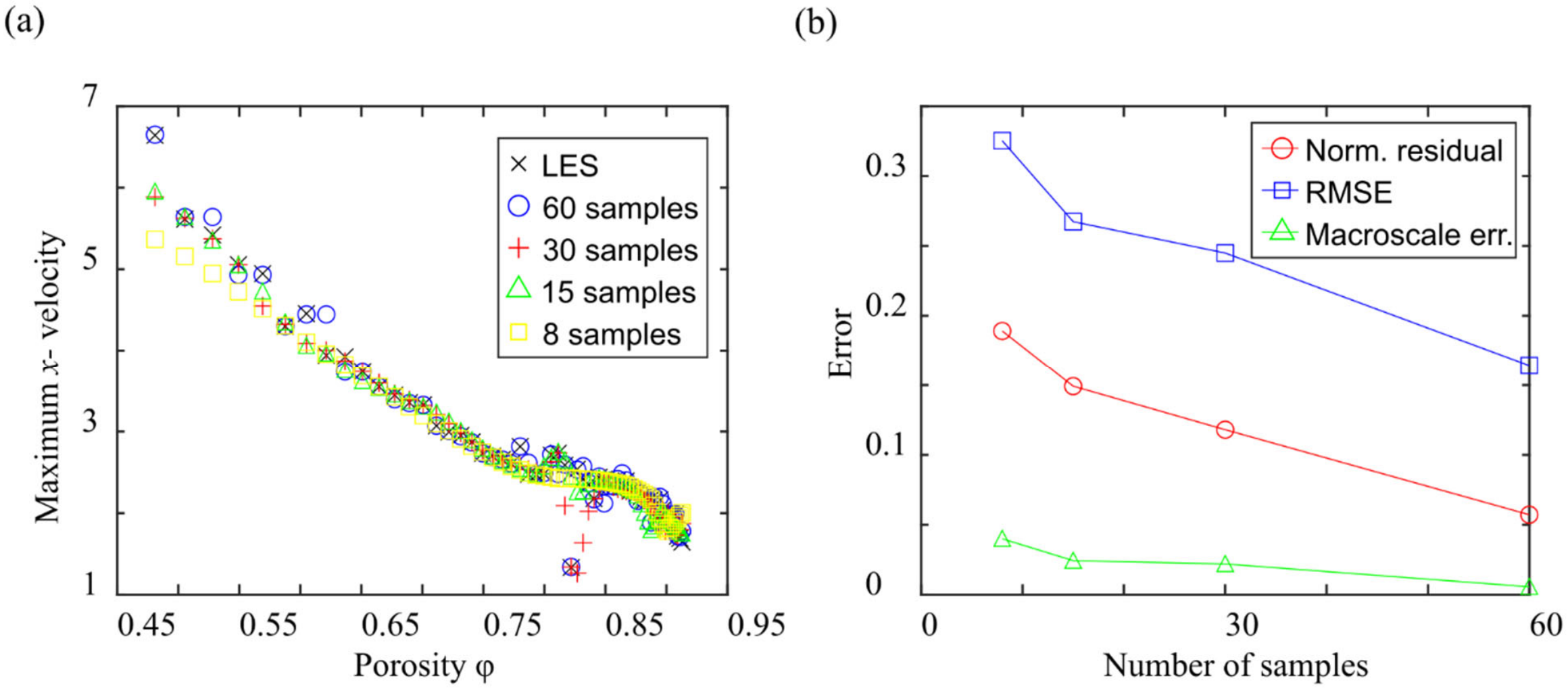

**Fig. 5** (a) The maximum *x*- velocity inside the fluid domain versus the porosity for LES and CNN models. (b) The variation of modeling error with the number of samples used to train the CNN model.

## 4. CONCLUSIONS

The microscale turbulent flow field inside an infinitely periodic, homogeneous porous medium is modeled by using Convolutional Neural Networks. The CNN model is trained to predict the flow field when the porosity of the porous medium is varied in the range from 0.454 to 0.913 with less than 5% modeling error at the macroscale by using only 8 samples obtained from Large Eddy Simulation. High accuracy with a small number of training samples will enable a comprehensive CNN model that takes other sensitive parameters into consideration, such as the solid obstacle shape, arrangement, and the Reynolds number. The CNN model offers a speed up of $O(10^6)$ over LES that can be leveraged for real-time simulations of porous media flow. The normalized residual and the Root Mean Squared Error become satisfactory when 60 samples are used to train the neural network. It is noted that the high magnitude of the normalized residual and RMSE does not influence the prediction of the volume-averaged turbulence variables commensurately. The mismatch in the position of the vortex shear layer and the gradient in the boundary layer are the main sources of error in the CNN model. While the vortex shear layer position is not crucial to macroscale modeling, the boundary layer prediction needs to be accurate for the drag terms that appear in the governing equations. To overcome this, the CNN architecture can be modified to predict surface quantities independently, or the training velocity function can be made smooth at the solid obstacle surface by Lagrange polynomial reconstruction. It is also noted that the CNN model is able to predict the flow separation and stagnation points accurately, which is of interest to the aerodynamics community. The CNN model has been shown to be effective in predicting the microscale flow field for the *x*- velocity for a range of porosities. There is an implicit understanding that this ability will be translated to other flow variables, such as the turbulence kinetic energy and the turbulence dissipation rate. However, this will need to be demonstrated in future work. The model error characterization for other geometric and flow parameters, and a more comprehensive turbulence model development should also be undertaken in future work.

## ACKNOWLEDGMENT

The authors acknowledge with gratitude the support of the National Science Foundation (Award No. CBET-2042834).

## NOMENCLATURE

| | | | | | |
|---|---|---|---|---|---|
| $d$ | solid obstacle diameter | ( m ) | $g_i$ | applied pressure gradient | ( $ms^{-2}$ ) |
| $s$ | pore size | ( m ) | $Re_p$ | pore-scale Reynolds number | ( - ) |
| $\varphi$ | porosity | ( - ) | $u_m$ | filtration velocity | ( $ms^{-1}$ ) |
| $f_i$ | IBM forcing term | ( $ms^{-2}$ ) | SDF | signed distance function | ( m ) |